# Real-Time Confidence Detection through Facial Expressions and Hand Gestures

Tanjil Hasan Sakib
*Department of Computer Science*
*American International University-Bangladesh*
Dhaka, Bangladesh
20-43633-2@student.aiub.edu

Samia Jahan Mojumder
*Department of Computer Science*
*American International University-Bangladesh*
Dhaka, Bangladesh
20-43474-1@student.aiub.edu

Rajan Das Gupta
*Department of Computer Science*
*American International University-Bangladesh*
Dhaka, Bangladesh
18-36304-1@student.aiub.edu

Md Imrul Hasan Showmick
*Department of Computer Science*
*Brac University*
Dhaka, Bangladesh
imrul.hasan.showmick@gmail.com

Md. Yeasin Rahat
*Department of Computer Science*
*American International University-Bangladesh*
Dhaka, Bangladesh
20-43097-1@student.aiub.edu

Md. Jakir Hossen
*Department of Computer Science*
*Multimedia University*
Malaysia
jakir.hossen@mmu.edu.my

*Abstract*— Real-time face orientation recognition is a cutting-edge technology meant to track and analyze facial movements in virtual environments such as online interviews, remote meetings, and virtual classrooms. As the demand for virtual interactions grows, it be- comes increasingly important to measure participant engagement, attention, and overall interaction. This research presents a novel solution that leverages the Media Pipe Face Mesh framework to identify facial landmarks and extract geometric data for calculating Euler angles, which determine head orientation in real time. The system tracks 3D facial landmarks and uses this data to compute head movements with a focus on accuracy and responsiveness. By studying Euler angles, the system can identify a user's head orientation with an accuracy of 90%, even at a distance of up to four feet. This capability offers significant enhancements for monitoring user interaction, allowing for more immersive and interactive virtual experiences. The proposed method shows its reliability in evaluating participant attentiveness during online assessments and meetings. Its application goes beyond engagement analysis, potentially providing a means for improving the quality of virtual communication, fostering better understanding between participants, and ensuring a higher level of interaction in digital spaces. This study offers a basis for future developments in enhancing virtual user experiences by integrating real-time facial tracking technologies, paving the way for more adaptive and interactive web-based platform.

## I. INTRODUCTION

Facial gestures, including facial expressions, body posture, and hand movements, are crucial for human communication and are key indicators of confidence. Non-verbal communication constitutes a significant portion of interpersonal interactions, often providing deeper insights than verbal cues. Despite their importance, current systems typically focus on isolated behavioral signals and struggle to monitor real-time engagement, particularly in online environments like virtual classrooms and remote interviews.

Traditional methods lack the ability to track essential indicators such as gaze direction, lip movement, and hand gestures effectively. In response, researchers have explored automated approaches using computer vision and AI to monitor various facial behaviors. However, these approaches often fail to integrate multiple indicators to assess confidence comprehensively.

This study introduces a real-time confidence detection system that combines facial expressions, head orientation, hand gestures, and blink rates using MediaPipe's machine learning models. By applying a weighted scoring method, the system offers a dynamic and holistic evaluation of confidence, with applications in public speaking training, virtual interviews, remote education, and online proctoring. The approach aims to improve engagement monitoring, performance evaluation, and human-computer interaction in virtual settings.

## II. LITERATURE REVIEW

Real-time confidence detection has gained ground with the ad- vancement of machine learning and facial recognition technolo- gies. It is widely applied in areas like virtual conversations, online learning environments, and human-computer interaction (HCI). Researchers have explored several techniques and models to track facial expressions and hand gestures in real-time, adding to a deeper understanding of communication and confidence perception.

### A. Facial Gesture Analysis and Confidence Detection

Facial expression recognition systems, such as the one developed using Histogram of Oriented Gradients (HOG) features with multiclass SVMs and random regression trees, have shown high accuracy but suffer from limited real-time performance due to low frame rates [33, 55]. However, relying solely on facial expressions has proven insufficient for accurate emotion detection in human-computer interaction, as demonstrated by [52], leading to the adoption of multimodal approaches. These combine additional behavioral cues—like hand gestures, gaze direction, and head movements—to improve detection accuracy [15]. For instance, [28] explored confidence and engagement detection in virtual interviews using a combination of facial gestures, emotion recognition, and eye tracking. Their findings emphasized that integrating hand gestures with facial expressions significantly enhances the reliability of real-time confidence assessment systems [9][58].

### B. Multimodal Data for Confidence Detection

Multimodal data has proven essential in improving the accuracy of confidence detection systems. [4][56] proposed the FILTWAM framework, which integrates facial expressions and vocal data to detect emotions in e-learning environments, demonstrating notable accuracy improvements in real-time contexts compared to single-modality systems. In the domain of online learning, [38] explored the use of convolutional neural networks (CNNs) to assess student engagement and support adaptive teaching strategies. Although their system achieved moderate accuracy, it highlighted the ongoing challenges of developing high-accuracy, real-time confidence detection tools for virtual education.

*C. Hand Gesture and Head Pose Detection*

Hand gesture analysis plays a key role in identifying confidence levels, with [30][57] demonstrating that controlled, moderate gestures correlate with confidence, whereas erratic movements often indicate anxiety. When combined with facial cues, hand gesture data significantly enhances the reliability of real-time trust recognition systems [40][59]. Similarly, head pose detection has been shown to contribute meaningfully to confidence assessment. [38] and [32][60] found that head orientation—specifically yaw, pitch, and roll angles—serves as an indicator of attention and focus, where steady and intentional movements are commonly associated with higher confidence in both public speaking and virtual interactions.

## III. ARCHITECTURE

Figure 1 presents the architecture of the real-time confidence detec- tion system. The process begins with capturing a real-time video stream, followed by frame capture and landmark detection to iden- tify key facial and hand features.

Next, the facial landmark extraction stage analyzes face move- ment, gaze, blink rate, lip movement, mouth openness, and hand motion. These features are processed at 30ms per frame to compute an average confidence score, which is converted into a percentage for final output

## IV. RESEARCH METHODOLOGY

The creation Science Research Methodology (DSRM) is used to guide the creation, implementation, and evaluation of a real-time confidence detection system. DSRM is a structured method to problem-solving in research, focusing on creating innovative so- lutions rather than simply analyzing existing phenomena. Themethodology for this study followed the steps of problem defini- tion, goal setting, artifact design, and evaluation [22, 46]., while the system integrates facial gesture behaviors to compute a compre-hensive confidence score for each frame of video input.

*A. Conceptual Framework*

This project develops a system for real-time behavioral tracking in virtual environments, such as online interviews and remote meetings, addressing the need for accurate insights into user engagement, focus, and confidence. By utilizing MediaPipe's Face Mesh and Hands models, the system tracks facial landmarks and hand gestures to measure key behavioral indicators. It calculates confidence values based on features like gaze direction, blink rate, head pose, and hand movements. The system was tested with an average accuracy of 90% and integrated into web platforms for real-time monitoring. Future enhancements include adding voice analysis and multi-user tracking to broaden its application across various sectors.

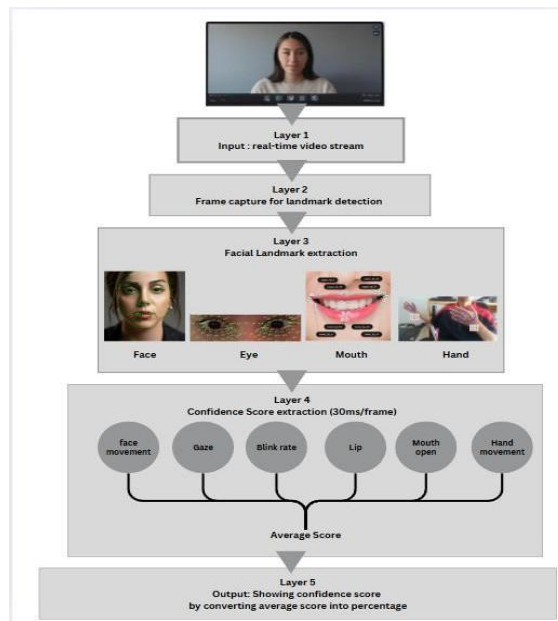

Fig. 1.System Architecture of the Real-Time Confidence Detection scoring.

*B. Conceptual Model: Design Science Research Methodology (DSRM)*

This study adopts the Design Science Research Methodology (DSRM) as its foundational framework, emphasizing not just the understanding of user behavior, but the purposeful design and evaluation of a real-time confidence detection system. The artifact developed in this research integrates facial and hand gestures to assess user confidence during virtual interactions—an area where traditional non-verbal cues like eye contact and body language are often lacking. Through the structured stages of DSRM, the study begins by identifying the challenge of confidence assessment in digital environments, followed by setting clear objectives for a system that can accurately and responsively interpret behavioral cues. The system is designed to process real-time image data, extracting features such as smiles, blinks, lip movements, and hand gestures at a rate of 30 milliseconds per frame. These features are classified and mapped to estimate the user's confidence level. The solution is then demonstrated in practical settings like online interviews and virtual classrooms to show its real-world applicability. Its effectiveness is evaluated by comparing the system's outputs with human judgments of confidence in controlled experiments. Finally, the study communicates its findings, highlighting both the system's performance and areas for future improvement. This research underscores how a design-driven approach can bridge the gap between human psychology and computational systems, offering timely, multimodal feedback to support more confident virtual communication.

*C. Data Collection*

Figure 3 shows the data collection process, where participants' facial expressions and hand movements were recorded via webcam during a 2-minute speech. The system computed live confidence scores per frame (30ms) and aggregated metrics like smile, blink rate, head pose, and gestures for analysis.

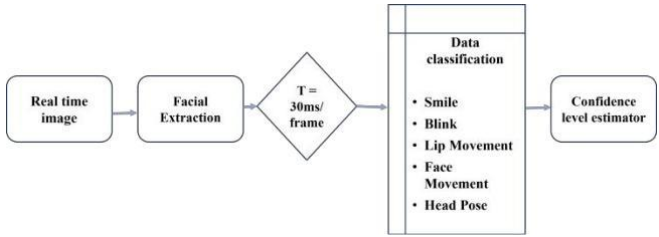

Fig. 2. A flowchart representing a proposed model for real-time image processing and confidence estimation.

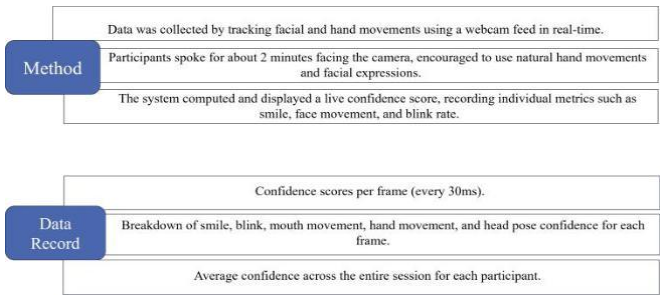

Fig. 3. Flowchart of the data collection process for real-time confidence scoring.

Data was collected from ten participants who delivered two-minute speeches while being recorded via webcam, maintaining natural facial expressions and hand gestures. The recorded videos served as the foundation for analyzing key behavioral indicators of confidence, including smile detection, blink rate, head movement, hand gestures, lip movement, and gaze steadiness. Machine learning techniques, such as Conventional Neural Networks and eye landmark detection, were applied to extract and interpret these features. Smiles and steady gazes were found to strongly correlate with higher confidence, whereas frequent blinking, erratic head movements, rapid hand gestures, prolonged speech pauses, and distracted gazes indicated lower confidence levels. Each video frame was analyzed in real-time to calculate a dynamic confidence score, combining these behavioral cues to continuously monitor fluctuations in participant confidence. The system offered a visual output of the confidence scores, providing valuable insights into real-time behavioral adaptation in virtual communication settings.

## V. Results and Analysis

This section presents the results of testing the real-time confidence detection system, which evaluates multiple facial gestures to generate a continuously updated confidence score. Using data collected during two-minute participant speeches, the system analyzed hand movements, facial expressions, blink rate, head pose, lip movement, and gaze steadiness. Moderate and smooth hand gestures, genuine smiles indicated by higher lip aspect ratios, steady head positioning, active lip movement during speech, and a focused gaze were all positively correlated with higher confidence levels. Conversely, excessive blinking, erratic head movements, prolonged lip stillness, and frequent gaze shifts were associated with lower confidence. By tracking these behavioral cues individually and combining them through a weighted average, the system provided a detailed, real-time assessment of participants' confidence throughout their interactions.

### A. Confidence Detection System

Hand gesture speed and smoothness were analyzed to assess confidence. Moderate movements between 0.2–0.5 m/s correlated with high confidence scores (0.9–1.2), while speeds above 0.5 m/s indicated nervousness and reduced scores (0.4–0.8). Overall, 70% of participants with controlled gestures scored above 0.9, 25% with slightly faster gestures (0.5–0.7 m/s) scored between 0.6 and 0.8, and 25% with rapid or erratic gestures scored below 0.5.

Smile detection was performed using the lip aspect ratio, identifying smiles when the ratio exceeded 1.5. Smiling increased confidence scores by a factor of 1.2, reinforcing associations with friendliness and engagement. Participants who smiled frequently scored between 0.9 and 1.2, while occasional smilers ranged from 0.6 to 0.8, and infrequent smilers scored below 0.6. Overall, 60% of participants who smiled regularly scored above 0.9, 25% with occasional smiles scored between 0.6 and 0.8, and 30% with rare smiles scored below 0.6. Although smiling strongly correlated with higher confidence, it was most effective when combined with other positive behaviors like moderate hand gestures and stable posture; isolated smiling with negative gestures still led to lower scores.

Blink rate was computed by measuring inter-blink intervals via eye landmarks. A blink rate exceeding 15 blinks per minute—ndicative of cognitive overload or stress—corresponded to a markedly lower confidence score (≈0.4). In contrast, participants with normal blink rates scored in the moderate (0.6–0.8) to high (0.9–1.0) confidence range. Those with moderately elevated blink rates (12–15 blinks/min) scored between 0.6 and 0.8, whereas those with excessive blinking scored below 0.5. Overall, 50% of participants with normal blink rates scored above 0.8, compared to 15% of the moderately elevated group (scoring 0.6–0.8) and 35% of the excessive-blinking group (scoring <0.5). These findings indicate that blink rate is a strong indicator of cognitive load and nervousness, effectively distinguishing confident participants from those under stress.

Head movement was analyzed by tracking yaw, pitch, and roll angles, where deviations beyond ±10° led to reduced confidence scores. Excessive or erratic head movements lowered confidence to around 0.4, reflecting signs of distraction or discomfort. Participants maintaining steady head postures generally scored in the medium (0.6–0.8) to high (0.9–1.0) confidence range, while those with moderate deviations scored between 0.6 and 0.8. Frequent or irregular head movements resulted in low confidence scores (<0.6). Overall, 55% of participants with stable posture scored above 0.8, 25% with moderate deviations scored between 0.6 and 0.8, and 40% with frequent movement scored below 0.6. These results highlight that steady head posture signals higher focus and confidence, although it must be supported by positive hand gestures and blink patterns for a complete assessment.

Lip movement during speech was monitored to assess confidence, where active lip activity indicated engagement, and prolonged stillness (over 5 seconds) suggested hesitation. Participants who maintained steady lip movements scored high confidence (0.9–1.2), while those with occasional pauses scored medium confidence (0.6–0.8), and participants with prolonged lip stillness scored low confidence (below 0.6). About 65% of participants showing regular lip movement scored above 0.9, 20% with occasional inactivity scored between 0.6 and 0.8, and 15% with prolonged stillness scored below 0.6. The analysis confirmed that consistent lip activity is a strong signal of confidence, although it is most effective when combined with other positive behaviors like smiling and maintaining a steady head posture.

The system evaluated gaze direction and stability to measure engagement, finding that a steady gaze correlated with higher confidence, while frequent gaze shifts suggested uncertainty. Participants who maintained a constant gaze scored high confidence (0.9–1.2), those with occasional gaze shifts scored medium confidence (0.6–0.8), and participants with frequent adjustments scored low confidence (below 0.5). Overall, 70% of participants with steady gaze scored above 0.9, 20% with occasional shifts scored between 0.6 and 0.8, and 10% with frequent shifts scored below 0.5. The analysis confirmed that gaze constancy is a critical indicator of confidence and engagement.

### B. Overall Confidence Score Results

By aggregating the results from all four facial gestures, the system calculated an overall confidence score for each participant, categorizing them into three levels. High confidence (0.9–1.2) was observed in participants who maintained steady postures, smiled frequently, and displayed moderate hand gestures, with stable blink rates and head movements reinforcing their scores. Medium confidence (0.6–0.8) was associated with minor signs of cognitive load, such as slightly elevated blink rates or occasional head movements, indicating mild discomfort without overt nervousness. Low confidence (0.4–0.5) appeared in participants who exhibited excessive blinking, erratic head movements, or rapid hand gestures, reflecting cognitive overload, apprehension, or anxiety.

### C. Weighting Calculation

The system first computes the confidence score for each individual facial gesture. These scores are then weighted according to the relative importance of each gesture before being combined into the total confidence score. For instance, if the gaze confidence score is 0.9 and its assigned weight is 15%, its contribution to the total score would be 0.9 × 0.15
= 0.135. This weighted approach ensures a balanced evaluation, allowing each gesture to influence the final confidence score proportionally. Figure 4 illustrates the process of weighting and integrating individual facial gesture scores into the overall confidence calculation.

### D. Data Summary

Here, the 60% of participants displayed high confidence, which was reflected in their consistent use of facial gestures indicating calmness and engagement. These participants demonstrated controlled behaviors such as moderate hand gestures, steady head posture, and frequent smiling. In contrast, 25% of participants exhibited medium confidence, showing subtle signs of nervousness or distraction, such as minor variations in blink rate or slight head movements. Finally, 15% of participants demonstrated low confidence, exhibiting clear signs of cognitive stress or discomfort, including excessive blinking, erratic hand movements, or frequent head shifts. These behaviors were associated with lower overall confidence scores.

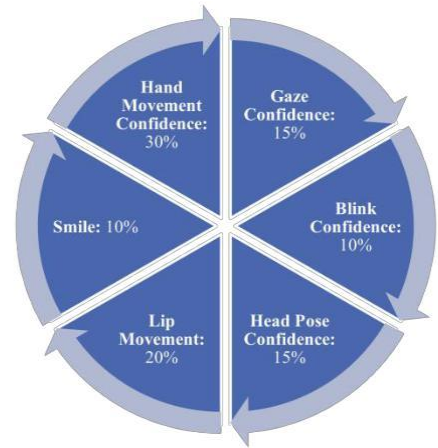

Fig. 4. A flowchart illustrating the process of weighting and combining individual confidence scores for real-time estimation.

### E. Weighting Calculation

The confidence detection system revealed important insights into the function of facial gestures in perceived confidence. The follow- ing main findings emerged:

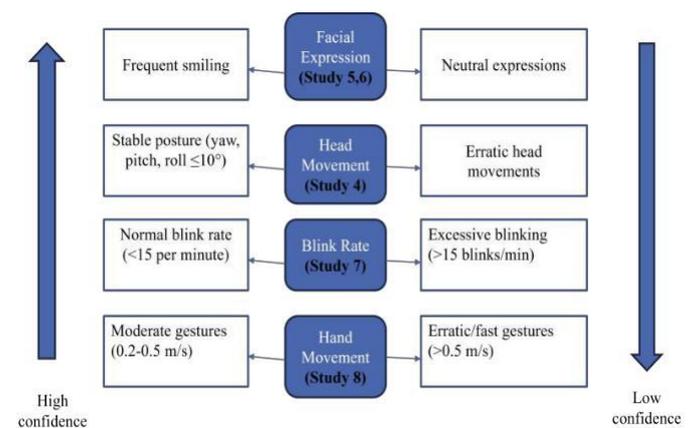

Fig. 5. Flowchart showing the weighting and aggregation of individual gesture scores to compute the confidence factor.

Figure 5 illustrates the process of calculating the confidence factor by assigning weights to individual facial gestures, such as expressions, head movements, blink rate, and hand movements. These weighted scores are then combined to determine the final confidence score.

*F. Correlation with Human Evaluation*

The data was collected during a 2-minute speaking session. Figure 6 shows a real-time confidence detection system using Mediapipe to track facial and hand landmarks, evaluating gestures, smiles, blink rates, and head movements, and providing a confidence score (e.g., 90.77%) for virtual interviews or e-learning. Figure 6 displays the system calculating a 90.00% confidence score for a participant who is actively engaged, smiling, and maintaining eye contact, which boosts the score.

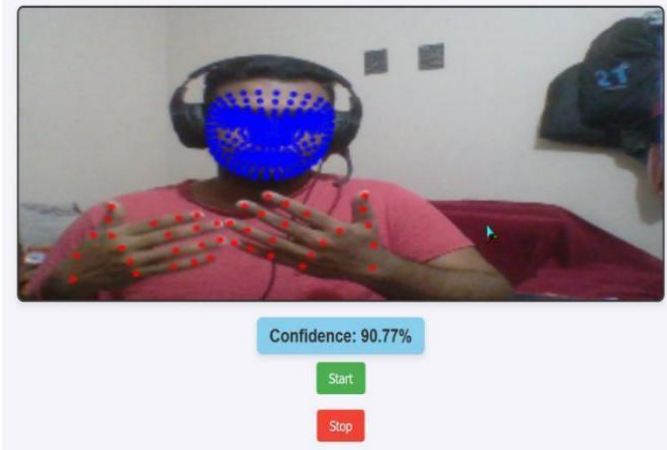

Fig. 6. Participant uses his hand while talking, showing confidence at 90.77%.

## VI. RESULTS AND ANALYSIS

This research introduces a real-time confidence detection system that analyzes non-verbal cues, such as facial expressions, gaze tracking, and hand gestures, to assess confidence instantly. It offers valuable insights for communication training, public speaking, and interview preparation. However, improvements could enhance the system's capabilities. Integrating voice analysis could refine confidence assessment by including speech patterns, tone, and intonation. Multi-face detection would allow the system to evaluate confidence in group settings, while advanced gesture and eyebrow tracking could improve emotional analysis. Real-time visual feedback and confidence breakdowns could further enhance user experience, and integration with VR/AR environments would support immersive training. Additionally, cross-cultural adaptation and personalized feedback would increase accuracy by considering regional body language variations.

| Factor | Weight (%) |
|---|---|
| Hand Gestures | 30% |
| Facial Expressions (Smile) | 10% |
| Lip Movement | 10% |
| Blink Rate | 10% |
| Head Movement | 15% |
| Gaze Confidence | 10% |

*Table 1: Contribution of Different Gestures to Confidence Score*

## ACKNOWLEDGEMENT

We would like to thank Multimedia University and ELITE Lab for supporting this research.